\documentclass[aps,prl,twocolumn,superscriptaddress,amsfonts,amsmath,amssymb]{revtex4-1}
\pdfoutput=1
\usepackage{amsmath}
\usepackage{amssymb}
\usepackage{graphicx}
\usepackage{xcolor}
\usepackage[plainpages=false, pdfpagelabels, bookmarksopen, pdfborder={0 0 0},colorlinks=true]{hyperref}

\DeclareMathOperator{\Tr}{Tr}

\begin{document}
\title{Decoherence and Decay of Two-level Systems due to Non-equilibrium Quasiparticles}
\author{Sebastian Zanker}
\affiliation{Institut f\"ur Theoretische Festk\"orperphysik, Karlsruhe Institute of Technology, D-76128 Karlsruhe,
Germany}
\author{Michael Marthaler}
\affiliation{Institut f\"ur Theoretische Festk\"orperphysik, Karlsruhe Institute of Technology, D-76128 Karlsruhe,
Germany}
\author{Gerd Sch\"on}
\affiliation{Institut f\"ur Theoretische Festk\"orperphysik, Karlsruhe Institute of Technology, D-76128 Karlsruhe,
Germany}
\begin{abstract}

  It is frequently observed that even at very low temperatures the number of quasiparticles in superconducting materials is higher than predicted by standard BCS-theory. These quasiparticles can interact with two-level systems, such as superconducting qubits or two-level systems (TLS) in the amorphous oxide layer of a Josephson junction. This interaction leads to decay and decoherence of the TLS, with specific results, such as the time dependence, depending on the distribution of quasiparticles and the form of the interaction. We study the resulting decay laws for different experimentally relevant protocols.
\end{abstract}
\maketitle
\section{Introduction}
Superconducting quantum devices have a wide range of applications. Due to the weak dissipation in the superconducting state they are promising candidates for building large scale quantum information systems \cite{testname,Devoret08032013}. 
They are easily controlled and measured by electromagnetic fields, but for the same reason they couple rather strongly to the environment and are prone to decoherence \cite{Shnirman2002}. Much effort has been put into understanding and minimizing various noise sources.

One ubiquitous source of decoherence arises from two level systems (TLS), such as bistable defects residing in dielectric substrates, disordered interfaces, surface oxides, or inside the barriers of Josephson junctions \cite{QubitTLSCoupling}.
While originally introduced to explain anomalous properties of glasses at low temperatures \cite{TSinAS} 
there is evidence that they are also an important source of decoherence for superconducting qubits \cite{testname} or superconducting resonators \cite{Faoro}.
A bath of TLS can explain the $1/f$ noise which limits the performance of many devices \cite{DecoherenceFromTLS}, while fluctuations of very slow TLS induce long-time parameter shifts. Still, the microscopic origin of those TLS remains unclear. Some potential sources are small groups of atoms that tunnel between two stable positions, or dangling bonds, or hydrogen defects. A better experimental as well as theoretical understanding of those TLS has been the focus of much recent work
\cite{ColeAtomic,PhillipsReview,TunnelingHydrogen,LossesMultilayer,PhysRevLett.95.210503}. 

Recent experiments demonstrated the coherent control of TLS, residing inside the amorphous layer of a phase qubit's Josephson junction, with help of this qubit \cite{CoherentControl}. It was possible to carry out typical coherence experiments as used in magnetic resonance or other qubit experiments \cite{Lisenfeld}. 
The aim of those experiments is a better understanding of the microscopic nature of individual TLS as well as their respective environment responsible for TLS state fluctuations, with the ultimate goal to reduce their detrimental effects.

In this paper we analyze the decoherence of charged TLS residing inside the amorphous layer of a Josephson junction \cite{QubitTLSCoupling, Grabovskij12102012} due to scattering and tunneling of  non-equilibrium quasiparticles in the superconducting leads. 
Experiments suggest that even at low temperatures the number of quasiparticles is large as compared to the predictions of equilibrium BCS theory \cite{PhysRevLett.103.097002}. Similar as observed for superconducting qubits \cite{PhysRevB.85.144503} or the dynamics of Andreev bound states \cite{Olivares14} the scattering with quasiparticles provides an intrinsic noise source also for the TLS. 
We investigate the TLS decoherence properties due to the coupling between the TLS and the quasiparticles. We find characteristic differences for quasiparticles which scatter back to the same superconducting electrode and those which tunnel across the junction. The effect of the latter depends on the phase difference. Our results apply both for the TLS, which are the main focus of the present paper, as well as for qubit decoherence \cite{Zanker15}. In fact for the latter the effect of tunneling electrons is more pronounced.

\section{The Model}
We consider the scattering of quasiparticles from a two-level system located inside the amorphous barrier of an aluminum oxide Josephson junction.
The model  Hamiltonian reads as
\begin{equation}\label{eq:modelHamiltonian}
 H = H_{\rm TLS} +H_{\rm qp} + H_C,
\end{equation}
where $H_{\rm TLS}$ is the TLS Hamiltonian, $H_{\rm qp}$ are the free quasiparticle Hamiltonians of the left and right lead and $H_C$ is the coupling between both subsystems.
Since the microscopic nature of TLS remains unclear, we use the phenomenological TLS standard model for $H_{\rm TLS}$ of Ref.~\cite{PhilipsSTM}. It describes the TLS as an effective charged particle trapped in a double well potential with asymmetry $\epsilon$ between the two potential minima and tunneling amplitude $\Delta_0$,
\begin{equation}
 H_{\rm TLS}=\frac{\epsilon}{2}\sigma_z+\frac{\Delta_0}{2}\sigma_x.
\end{equation}
In the TLS eigenbasis the Hamiltonian reduces to $H_{\rm TLS} = \frac{1}{2}E_{\rm TLS}\,\tilde\sigma_z$
with  $E_{\rm TLS}=\sqrt{\Delta_0^2+\epsilon^2}$. Electrons in nearby leads couple to the TLS' electric dipole moment and induce an interaction that is well established in the context oft metallic glasses \cite{BlackInteraction}:
\begin{equation}\label{eq:Coupling}
 H_C = \sigma_z \hat V = \sigma_z \sum_{kk'}\left(g_{kk'}c_{k}^{\dagger}c_{k'}^{\phantom\dagger}+h.c.\right) .
\end{equation}
Here, $g_{kk'}$ is the coupling strength between electrons and the TLS dipole and $c_k$ is an electron annihilation operator with multi-index $k=\{\vec k,\sigma,\alpha\}$ that includes electron momentum $\vec k$, spin $\sigma$, and the index $\alpha=l,r$ of the lead (\emph{left} or \emph{right}), where the electron resides.
In general we can distinguish two processes: Electrons that tunnel through the junction ($\alpha\neq\alpha'$) while interacting with the TLS and electrons, that scatter back into their original lead ($\alpha=\alpha'$).
Because of the exponentially decaying electron wave function inside the junction and the localized character of the TLS wave function, the interaction rapidly decreases for TLS  away from the junction edges.
We expect that for most TLS scattering electrons are the main source of TLS decoherence, while the influence of tunneling electrons is insignificant. This is quite different for the decoherence of a qubit, where only tunneling electrons couple to the qubit and induce decoherence.
Furthermore, electrons that contribute to TLS decoherence have energies close to the Fermi energy with momentum $|\vec k|=k_F$. 
Hence, we take the direction average and introduce the direction-averaged coupling constant $g_{\alpha\alpha'}^2\equiv\langle g_{kk'}^2\rangle$. Since $|\vec k|\approx k_F$ the averaged coupling constant does not depend on energy.

The free particles in the leads are  Bogoliubov quasiparticles with mixed electron- and hole-like nature and creation operators $a_{k}^{\dagger}$  and Hamiltonian
\begin{equation}
 H_{\rm qp} = \sum_{k}E_{k}a_{k}^{\dagger}a_{k}^{\phantom\dagger}
\end{equation}
Their energy is $E_k=\sqrt{\xi_k^2+\Delta_{\rm bcs}^2}$ with $\xi_k$ being the electron energy in the normal state. Rewriting the coupling Eq. \eqref{eq:Coupling} in terms of quasiparticle operators we find
\begin{equation}\label{eq:V}
 \hat V = \sum_{kk'}g_{\alpha\alpha'}\left(e^{i\varphi/2}u_{k}u_{k'}-e^{-i\varphi/2}v_{k}v_{k'}\right)a_k^\dagger a_{k'}^{\phantom\dagger}+h.c.
\end{equation}
with coherence factors $u_k^2=1-v_k^2=\frac12 (1+\xi_k/E_k)$. For tunneling quasiparticles $\varphi$ is the superconducting phase difference across the junction. The phase difference $\varphi$ vanishes for scattering quasiparticles.

An important quantity in the context of decoherence is the noise spectral density 
\begin{equation}\label{eq:spectralDensityDefinition}
  S_{V}(\omega)=\int\frac{dt}{2\pi}\langle \hat V(t)\hat V(0)\rangle e^{i\omega t}
 \end{equation}
where the average is over the quasiparticle states. With Eq. \eqref{eq:V} we find
\begin{eqnarray}\notag
 S_V(\omega) = & 4N_0^2g_{\alpha\alpha'}^2\int_{\Delta_{\rm bcs}}^\infty\int_{\Delta_{\rm bcs}}^\infty dEdE'\,\rho(E)\rho(E')
 \\ \notag
 & \times \left(1- \frac{\Delta_{\rm BCS}^2}{EE'}\cos\varphi\right)
 \\ \notag
& \times \{f_\alpha(E)[1-f_{\alpha'}(E')]\delta(E-E'+\omega)
  \\ 
&+f_{\alpha'}(E')[1-f_\alpha(E)]\delta(E'-E+\omega)\} \label{eq:spectralDensity}
\end{eqnarray}
with the density of states at the Fermi energy of the normal state $N_0$, the BCS density of states $\rho(E)=E/\sqrt{E^2-\Delta_{\rm BCS}^2}$ and the quasiparticle distribution function $f(E_k)=\langle a_k^\dagger a_k^{\phantom\dagger}\rangle$. 
We assume that the quasiparticles can be described with equilibrium BCS gap and density of states, but with a non-equilibrium distribution function $f(E_k)$. 
Due to the square root singularity of the BCS density of states, the noise spectral density is log-divergent at low frequencies for \emph{tunneling} quasiparticles, e.g. $\varphi\neq0$. \emph{Scattering} quasiparticles, e.g. $\varphi=0$ have a finite spectral density at low frequencies.


\section{TLS Decoherence}
The time evolution of the TLS in the presence of the quasiparticle reservoirs is best described with the help of the reduced density matrix
\begin{equation}
 \rho(t) = \Tr\left[\varrho(t)\right]_{\rm qp} \equiv \begin{pmatrix}\rho_0&\rho_{01}\\\rho_{10}&\rho_1\end{pmatrix}
\end{equation}
that is obtained from the full density matrix after tracing out quasiparticle degrees of freedom. It evolves according to
\begin{equation}\label{eq:Dyson}
 \rho(t) = e^{-iH_{\rm TLS}t}\Tr_{\rm qp}\left[U_I^{\phantom\dagger}(t)\varrho(0)U_I^\dagger(t)\right]e^{iH_{\rm TLS}t}
\end{equation}
where $U_I(t,t_0)$ is the time evolution operator 
\begin{equation}
 U_I(t,t_0) = \text{Texp}\left[-i\int_{t_0}^tdt'\,H_{C,I}(t')\right],
\end{equation}
and $H_{C,I}(t')$ the coupling in the interaction picture. It is assumed that the initial density matrix,
$\varrho(t_0)\equiv\rho(t_0)\rho_{\rm qp}(t_0)$,  factorizes into a quasiparticle and TLS component and that initial correlations are irrelevant on experimental time scales.
We can distinguish two effects due to the quasiparticles: Decay and decoherence. The former describes exponential decay of diagonal elements of the TLS density matrix to their stationary state values, while the latter concerns the decay of off-diagonal elements. 

Transforming the coupling \eqref{eq:Coupling} into the TLS energy basis we find two contributions $\sigma_z\to \epsilon/E_{\rm TLS}\, \tilde\sigma_z+\Delta_0/E_{\rm TLS}\, \tilde\sigma_x$. 
The off-diagonal term $\sim \tilde\sigma_x$ induces transitions and is responsible for the decay rate $\Gamma_1$. It can be calculated in first-order perturbation theory \cite{PhysRevB.72.134519},
\begin{equation}\label{eq:gamma1}
  \Gamma_1 = \frac{\Delta_0^2}{E_{\rm TLS}^2}\left[S_V(E_{\rm TLS})+S_V(-E_{\rm TLS})\right].
\end{equation}
The diagonal coupling $\sim\tilde\sigma_z$ generates pure dephasing, determined by the low-energy part of the spectral density. Due to the strong energy dependence in this energy range pure dephasing does not lead to a simple exponential decay law. Rather the off-diagonal elements of the density matrix take the form
\begin{equation}\label{eq:dephasingRho}
 \rho_{\rm 10/01}(t) = e^{\pm i E_{\rm TLS}t}e^{-\frac12\Gamma_1 t}e^{-h(t)}
\end{equation}
where $h(t)$ describes the deviations from the simple exponential decay. It reduces to a linear time-dependence  only for flat spectral densities and long times.
This form of the density matrix follows from Eq. \eqref{eq:Dyson} and the fact that the TLS--quasiparticle coupling is diagonal in TLS space for pure dephasing. Due to the simple coupling we can pull all TLS operators through the trace and arrive at the form for the TLS density matrix given in \eqref{eq:dephasingRho} with
\begin{equation}\label{eq:F}
e^{-h(t)}\equiv\Tr_{\rm qp}\left[T_C\exp\left\{i\int_{t_0}
^t\hat V(t')dt'\right\}\rho_{\rm qp}(t_0)\right] \, .
\end{equation}
The contour time-ordering operator $T_c$ orders along a contour from $t_0$ to $t$ and back again. To further evaluate that expression we expand the exponential and introduce an additional approximation \cite{Zanker15}: 
We assume that we can split averages over $V_i$ operators in the form
\begin{equation}
\langle V(t_1)V(t_2)\dots V(t_n)\rangle = \prod_{\rm perm}\langle V(t_i)V(t_j)\rangle\cdots\langle V(t_k)V(t_l)\rangle
\end{equation}
With this approximation the quasiparticles behave similar to a Gaussian noise source \cite{PhysRevB.72.134519} and we find
\begin{equation}\label{eq:Ramsey}
h_R(t) = t^2\int
d\omega \, S_{qp}(\omega)\,\frac{\sin^2\left(\omega t/2\right)}{\left(\omega t/2\right)^2}
\end{equation}

This specific form for pure dephasing is well established in the context of magnetic resonance  or qubit experiments in a Ramsey protocol. 
The weighting function $g(\omega t) = \sin^2(\omega t/2)/(\omega t/2)^2$ has a pronounced peak for zero energy and decreases rapidly for larger $\omega$. Therefore, the Ramsey-type experiments are sensitive to the spectral density at low energies. 

Within the Gaussian approximation we can extend our analysis to more sophisticated measurement protocols, such as spin echo  or  more complicated refocusing techniques that suppress low-frequency noise contributions.
The dephasing function $h(t)$ for those protocols looks very much like the Ramsey function but with different filter functions depending on the particular pulse protocol \cite{nphys1994}
\begin{equation}\label{eq:GeneralDephasing}
 h(t) = t^2\int
d\omega \, S_{qp}(\omega)\,g(\omega t).
\end{equation}
E.g. for spin echo, which is the 'first order' improvement to the Ramsey experiment, the filter function is
\begin{equation}
  g_e(\omega t) = \sin^4\left(\omega t/4\right)/\left(\omega t/4\right)^2
\end{equation}
with a maximum slightly shifted to higher frequencies. For typical experimental times in the range of microseconds the filter function measures the spectral density at energy equivalents of several MHz.

\section{Non--Equilibrium Quasiparticles}\label{sec:Decoherence}
Based on the dephasing functions  \eqref{eq:GeneralDephasing} and \eqref{eq:Ramsey} as well as the decay rate \eqref{eq:gamma1} we are ready to analyze the dephasing process due to quasiparticles. The spectral density \eqref{eq:spectralDensity} depends on the quasiparticle distribution function.
Several experiments provide evidence that, even at low temperatures where quasiparticles should be exponentially suppressed with the BCS gap, finite densities of quasiparticles remain, estimated to be $n_{qp}\sim 10^{-6} \cdot\Delta_{\rm BCS}N_0$ \cite{PhysRevLett.103.097002}. 
Similar to the treatment of non-equilibrium quasiparticles for qubit decoherence, e.g. in Ref.~\cite{PhysRevB.86.184514}, we assume that both, the BCS gap and the density of states are not changed, but the distribution function is of a non-equilibrium form.
Although the exact form depends on experimental details most of the non-equilibrium quasiparticles have energies close to the superconducting gap because of scattering with phonons and among each other. 
We therefore assume that the distribution function has a width $\delta$ above the gap, which for a Fermi distribution is determined by temperature, $\delta_{\rm eq}\sim k_BT$, but here it is treated as a parameter. In the following we derive analytical forms for the different rates in the experimental relevant long-time limit $t\gg \delta^{-1}$.

\subsection{Decay}
The relevant energy scale for TLS decay is the TLS energy splitting $E_{\rm TLS}$ as evident from Eq. \eqref{eq:gamma1}. 
The TLS which can be probed by a qubit have energy splittings close to that of the qubit and thus fulfill $E_{\rm TLS}\gg \delta$. In order to evaluate the spectral density in this limit we introduce the normalized quasiparticle density
\begin{equation}\label{eq:xqp}
 x_{qp}=\frac{1}{\Delta_{\rm BCS}}\int_{\Delta_{\rm bcs}}^\infty dE\, \rho(E)f(E).
\end{equation}
For typical TLS energies $\Delta_{\rm BCS}\gg E_{\rm TLS}\gg\delta$ we can apply the 'low-energy' approximation to evaluate the spectral density at the TLS energy \cite{PhysRevB.86.184514, PhysRevB.85.144503}. In this limit all quasiparticle energies in Eq. \eqref{eq:spectralDensity} can be set to $\Delta_{\rm BCS}$. The only exception is the quasiparticle energy in the divergent BCS density of states together with the corresponding distribution function $f(E)$. They enter in the quasiparticle density \eqref{eq:xqp} and the spectral density, describing the decay, reads as
\begin{align}\notag
 S_V(E_{\rm TLS})& = 4N_0^2g_{\alpha\alpha'}^2\Delta_{\rm BCS}(x_{\rm qp,\alpha}+x_{\rm qp,\alpha'})\\
 &\times \rho(E_{\rm TLS}+\Delta_{\rm BCS})
\left(1-\frac{\Delta_{\rm BCS}}{\Delta_{\rm BCS}+E_{\rm TLS}}\cos\varphi\right) \, ,
\end{align}
while $\sim S(-E_{\rm TLS})$ and the resulting excitation rate is much smaller. This form for the high-energy spectral density and thus decay rate $\Gamma_1$ is well established in the context of qubit decay due to quasiparticles \cite{PhysRevB.84.064517,PhysRevLett.103.097002}.

\subsection{Ramsey and Spin Echo Dephasing}
To calculate the Ramsey dephasing rate \eqref{eq:Ramsey} we need an approximation for the low-energy spectral density. We proceed as in Ref.~\cite{Zanker15} and split the spectral density into a regular and a divergent part. For low energies, the former is flat and can be considered constant, while the latter is log divergent, $S_{\rm div}\sim(1-\cos\varphi)\log(\delta/\omega)$. We find the Ramsey dephasing function
\begin{align}\notag
 h_R(t) = \Gamma_2^*\,t+\pi N_0^2g_{\alpha\alpha'}^2[f_\alpha(\Delta_{\rm BCS})+f_{\alpha'}(\Delta_{\rm BCS})]
 \\ \label{eq:RamseyAna}
 \times[1-\cos(\varphi)]
\left[\gamma_e-1+\log(4\delta\cdot t)\right]t
\end{align}
with the Euler constant $\gamma_e$ and the pure dephasing rate
\begin{equation}
 \Gamma_2^*= 8\pi N_0^2g_{\alpha\alpha'}^2\int_{\Delta_{\rm BCS}}^\infty [f_\alpha(E)+f_{\alpha'}(E)]dE \, .
\end{equation}
For scattering quasiparticles we have $\cos\varphi=1$ and the divergent contribution vanishes. Thus, scattering particles induce simple exponential dephasing $\sim e^{-\Gamma_2^*t}$ with rate $\Gamma_2^*$. On the other hand, the dephasing effect of tunneling quasiparticles is dominated by the second term in \eqref{eq:RamseyAna} stemming from the divergent contribution.

The spin echo protocol filters out low energies, but we observe that the relevant energy scales are still much smaller than the width of the quasiparticles. 
Thus the calculation proceeds similar to the calculation for Ramsey dephasing, with the result
\begin{align}\notag
 h_e(t) = \Gamma_2^*\,t+\pi N_0^2g_{\alpha\alpha'}^2[f_\alpha(\Delta_{\rm BCS})+f_{\alpha'}(\Delta_{\rm BCS})]
 \\
 \times(1-\cos(\varphi)) \frac12\left[\gamma_e-1+\log(\delta \cdot t)\right]t.
\end{align}
Since the non-divergent part of the spectral density is almost flat for the relevant frequency scales, spin echo does not improve coherence and, similar to white-noise-induced dephasing, the pure dephasing rate $\Gamma_2^*$ is the same for both protocols, spin echo and Ramsey. Thus, for scattering electrons there is no measurable difference between both experimental protocols.aeternus20!0

On the other hand, for tunneling quasiparticles the second term comes into play and dominates dephasing. In this limit the ratio between spin echo and Ramsey is
\begin{equation}
 \lim_{t\to \infty}\frac{h_{e}(t)}{h_r(t)}=\frac12.
\end{equation}
This improvement in dephasing time due to spin echo is typical for noise with divergent spectral density at small energies and could be measured in an experiment.

\section{Conclusion}
We analyzed the decoherence of TLS located in disordered systems in vicinity to superconducting leads, especially inside the amorphous layer of a Josephson junction. We distinguish in our analysis between scattering quasiparticles, that cause decoherence for all the TLS mentioned above, and quasiparticles that tunnel through the junction.
If there exists a phase difference between the superconducting electrodes, the latter exhibit a log-divergent spectral density for low energies leading to increased and time-dependent dephasing rates
while the difference is negligible for the TLS decay. 
We further showed that the spin echo technique reduces the TLS decoherence rate due to tunneling particles, while it has little effect for scattering quasiparticles.
The results obtained for tunneling quasiparticles, arising form a divergent spectral density, apply to single-junction qubits, and they are sensitive to refocusing techniques. This opens possibilities to analyze the quasiparticle-environment of a qubit.

\section*{Acknowledgments}
We thank A. Bilmes, J. Lisenfeld and A. Shnirman for many useful discussions during the work on this paper. 
This work was supported by the German-Israeli Foundation for Scientific Research and Development (GIF).

\bibliography{bibTLSnoURLnoMonth}

\end{document}